\documentclass[twocolumn]{jpsj3}
\usepackage{txfonts}
\title{Critical Exponent of the Anderson Transition using Massively Parallel Supercomputing}

\author{Keith Slevin$^1$\thanks{slevin@phys.sci.osaka-u.ac.jp} and Tomi Ohtsuki$^2$}
\inst{$^1$Department of Physics, Graduate School of Science, Osaka University, 1-1 Machikaneyama, Toyonaka, Osaka 560-0043, Japan \\
$^2$Division of Physics, Sophia University, Chiyoda-ku, Tokyo 102-8554, Japan  }

\abst{To date the most precise estimations of the critical exponent for the Anderson transition have been made using the transfer matrix method.
This method involves the simulation of extremely long quasi one-dimensional systems. The method is inherently serial and is not well suited to modern massively parallel supercomputers. The obvious alternative is to simulate a large ensemble of hypercubic systems and average. While this permits taking full advantage of both OpenMP and MPI on massively parallel supercomputers, a straight forward implementation results in data that does not scale. We show that this problem can be avoided by generating random sets of orthogonal initial vectors with an appropriate stationary probability distribution. We have applied this method to the Anderson transition in the three-dimensional orthogonal universality class and been able to increase the largest $L\times L$ cross section simulated from $L=24$ (New J. Physics, {\bf 16}, 015012 (2014))
to $L=64$ here. This permits an estimation of the critical exponent with improved precision and without the necessity of introducing an irrelevant scaling variable. In addition, this approach is better suited to simulations with correlated random potentials such as is needed in quantum Hall or cold atom systems.}

\begin{document}
\maketitle

\section{Introduction}

The transfer matrix method was first introduced into the field of Anderson localisation by Pichard and Sarma\cite{pichard81}, and MacKinnon and Kramer\cite{mackinnon81,MacKinnon83}.
These papers reported the first numerical evidence in favour of the scaling theory of localisation.\cite{abrahams79}
Since that pioneering work the method has been employed extensively and very successfully
\cite{Kramer93,slevin97,slevin99,slevin14,ueoka14,slevin16,asada02}
to
estimate with high precision the critical exponents for the standard Wigner-Dyson symmetry classes\cite{Wigner51,Dyson61,Dyson62}
in various dimensions and for the quantum Hall effect.\cite{Huckestein90,slevin09,amado11,obuse12}

However, the method is inherently serial and is not well suited to modern massively parallel supercomputers.
Here, we describe an adaptation of the method that is better suited to such computers.
We show that the critical exponent can be estimated correctly by simulating an ensemble of cubes rather than the single very long quasi-one dimensional systems in the serial method.
The key point is to make the initial matrix used in the method a random matrix that
is sampled from an appropriate stationary probability distribution.
We demonstrate the effectiveness of the method by applying it to estimate the critical exponent for
the Anderson transition in the three dimensional orthogonal universality class.

\section{Model and Simulation Method}

\subsection{Anderson's Model of Localisation}
The Hamiltonian for Anderson's model of localisation\cite{anderson58} may be written in the form
\begin{equation}
H = \sum\limits_i {{E_i}} \left| i \right\rangle \left\langle i \right| - V \sum\limits_{\left\langle {ij} \right\rangle } {\left| i \right\rangle \left\langle j \right|} \,.
\end{equation}
Here, $\left| i \right\rangle$ is a localised orbital on site $i$ of a three dimensional cubic lattice.
The first sum is over all sites on the lattice and the second sum is over pairs of nearest neighbours.
We measure all energies in units of the hopping energy $V$ between nearest neighbour orbitals (so that in what follows $V$ does not appear explicitly, i.e. $V=1$).
The energies $E_i$ of these orbitals are assumed to be
identically and independently distributed random variables with a uniform distribution
\begin{equation}
p\left( {{E_i}} \right) = \left\{ {\begin{array}{*{20}{c}}
{{1 \mathord{\left/
 {\vphantom {1 W}} \right.
 \kern-\nulldelimiterspace} W}}&{\left| {{E_i}} \right| \le {W \mathord{\left/
 {\vphantom {W 2}} \right.
 \kern-\nulldelimiterspace} 2}}\,,\\
0&{{\rm{otherwise}}}\,.
\end{array}} \right.
\end{equation}
The parameter $W$ determines the degree of disorder.
Since this Hamiltonian commutes with the complex conjugation operator, i.e. a time reversal operator that squares to plus unity, this model is
in the orthogonal symmetry class.\cite{Wigner51,Dyson61,Dyson62}

\subsection{The Transfer Matrix}
We consider a system with a uniform square cross section $L \times L$ which we divide into layers labelled by their $x$ coordinate.
We then re-write the time independent Schr\"odinger equation for a state vector $\left| \Psi \right>$ and energy $E$
\begin{equation}
H\left| \Psi  \right\rangle  = E\left| \Psi  \right\rangle \,,
\end{equation}
in the following form
\begin{equation}
\left( {\begin{array}{*{20}{c}}
{{\psi _{x + 1}}}\\
{ - {\psi _x}}
\end{array}} \right) = {M_x}\left( {\begin{array}{*{20}{c}}
{{\psi _x}}\\
{ - {\psi _{x - 1}}}
\end{array}} \right) \,.
\end{equation}
Here, $\psi_x$ is the vector of wavefunction amplitudes on layer $x$ (in some suitable order)
\begin{equation}
 {\left( {{\psi _x}} \right)_{y,z}} = \left\langle {x,y,z\left| \Psi  \right\rangle } \right. \,.
\end{equation}
$H_x$ is the following sub-matrix of the Hamiltonian
\begin{equation}
  {\left( {{H_x}} \right)_{y,z,y',z'}} = \left\langle {x,y,z} \right|\left. H \right|\left. {x,y',z'} \right\rangle \,,
\end{equation}
and $M_x$ is the following $2N \times 2N$ transfer matrix (with $N=L^2$),
\begin{equation}\label{eq:TransferMatrix}
{M_x} = \left( {\begin{array}{*{20}{c}}
{{H_x} - E 1_N}&1_N\\
{ - 1_N}&0_N
\end{array}} \right) \,
\end{equation}
with $1_N$ and $0_N$ the $N \times N$ unit and zero matrices, respectively.
The boundary conditions in the transverse directions must also be specified.\cite{Slevin00} Throughout this paper, we set the energy at the band centre, i.e. $E=0$, and impose periodic boundary conditions in the transverse directions.

Since solutions of the time independent Schr\"odinger equation are stationary, the probability flux must be conserved by the
transfer matrix multiplication. This implies that the transfer matrix must satisfy the following relation
\begin{equation}\label{eq:CurrentConservation}
  {M_x^T}\Sigma_c M_x = \Sigma_c \,,
\end{equation}
where
\begin{equation}
  \Sigma_c  = \left( {\begin{array}{*{20}{c}}
0_N&{ - i_N}\\
i_N&0_N
\end{array}} \right) \,.
\end{equation}
Here, $i_N = i 1_N$.

\subsection{Serial Method}
We consider a very long quasi-one-dimensional bar of length $L_x$. The wave-function amplitudes on the first two layers are related to the wave-function amplitudes on the last two layers as follows
\begin{equation}
  \left( {\begin{array}{*{20}{c}}
{{\psi _{{L_x} + 1}}}\\
{ - {\psi _{L_x}}}
\end{array}} \right) = {M_{{L_x}}} \cdots {M_1}\left( {\begin{array}{*{20}{c}}
{{\psi _1}}\\
{ - {\psi _0}}
\end{array}} \right)\,.
\end{equation}
This involves the product of $L_x$ independently and identically distributed random matrices
\begin{equation}
  M = {M_{{L_x}}} \cdots {M_1}\,.
\end{equation}
According to the theorem of Oseledec,\cite{oseledec68} the following limiting matrix exists
\begin{equation}
  \Omega  = \mathop {\lim }\limits_{{{L_x}} \to \infty } \frac{{\ln {M^T }M}}{{2{{L_x}}}} \,.
\end{equation}
Note the limit depends on the particular sequence of random matrices not just on the distribution.
However, for the eigenvalues $\{ \gamma_i \}$ of $\Omega$ we obtain the same values for almost all sequences, i.e. with probability one.
These values are called Lyapunov exponents.
From Eq. (\ref{eq:CurrentConservation}) it can be shown that these eigenvalues occur in pairs of opposite sign. It is usual to number them as follows
\begin{equation}\label{eq:order}
  {\gamma _1} > {\gamma _2} >  \cdots {\gamma _N} >  \gamma_{N+1}=- {\gamma _N} >  \cdots  >\gamma_{2N}= - {\gamma _1} \,.
\end{equation}
To estimate the Lyapunov exponents we start with a $2N \times 2N$ orthogonal matrix, truncate the matrix product at
a very large but finite $L_x$, and perform a QR factorization of the result
\begin{equation}\label{eq:QR}
  QR = M {Q_0} \,.
\end{equation}
Here, $Q_0$ is a $2N\times 2N$ orthogonal matrix and $R$ is a $2N\times 2N$  upper triangular matrix  with positive diagonal elements.
We then define
\begin{equation}
  {\tilde \gamma _i} = \frac{1}{{L_x}}\ln {R_{i,i}} \,.
\end{equation}
In the limit of infinite length
\begin{equation}
  {\gamma _i} = \mathop {\lim }\limits_{{L_x} \to \infty } {\tilde \gamma _i} \,.
\end{equation}
For sufficiently large $L_x$, the $\left\{ {{{\tilde \gamma }_i}} \right\}$ may be used to estimate the Lyapunov exponents.
In practice, it is not possible to implement the calculation Eq. (\ref{eq:QR}) straightforwardly. Instead, to avoid round off error, it is necessary to
perform intermediate QR factorizations at regular intervals (say after every $q$ transfer matrix multiplications). The details are described in Ref. \citen{slevin14}.
Note that, for the purposes of finite size scaling, it is usual, though not essential,\cite{slevin01a} to focus on the smallest positive Lyapunov exponent $\gamma_N$.
The calculation of the negative Lyapunov exponents is then not necessary.
In this case, it is sufficient to make $Q_0$ a $2N \times N$ matrix with orthogonal columns, and
$R$ then becomes an $N\times N$  matrix.
This saves considerable computational time.

We next turn to the question of how to estimate the critical parameters.
The critical parameters of main interest are
the critical disorder $W_c$ separating the diffusive (metal) phase for $W<W_c$ from the localised (insulator) phase for $W>W_c$,
and the critical exponent $\nu$ that describes the divergence of the correlation (localisation) length $\xi$ at the critical point
\begin{equation}
  \xi \sim \left| W - W_c\right|^{-\nu} \,.
\end{equation}
These, and other critical parameters, are estimated by fitting the system size and disorder dependence of the dimensionless quantity
\begin{equation}\label{eq:Gamma}
  \Gamma =\gamma_N L\;,
\end{equation}
to a model derived from the following one-parameter scaling hypothesis
\begin{equation}\label{eq:ScalingGeneral}
  \left<
  \tilde{\gamma}_N
  \right>
  L_x
   =
  f\left(
  \frac{L_x}{\xi},\frac{L}{\xi },\frac{L}{\xi}
  \right) \,,
\end{equation}
with $f$ a scaling function.
For sufficiently long systems such that the effect of the initial matrix $Q_0$ becomes negligible,
we expect the right hand side of Eq. (\ref{eq:ScalingGeneral}) to be proportional to $L_x$.
Thus, after taking the limit of infinite sample length, Eq. (\ref{eq:ScalingGeneral}) reduces to
\begin{equation}
  \Gamma = \gamma_N L  = {f_{ \mathrm{Q1D} }}\left( {\frac{L}{\xi }} \right)\,,
\end{equation}
where an ensemble average or tilde are no longer needed and a related scaling function $f_{\mathrm{Q1D}}$
has been introduced.

This method requires the simulation of a single very long sample.
While this method has been employed very successfully in numerous simulations over the preceding decades, it is an inherently serial calculation that does not allow us to take advantage of modern parallel computers particularly those with hundreds of CPU nodes.

\subsection{Parallel Method}
\begin{figure}
\includegraphics[width=0.5\textwidth]{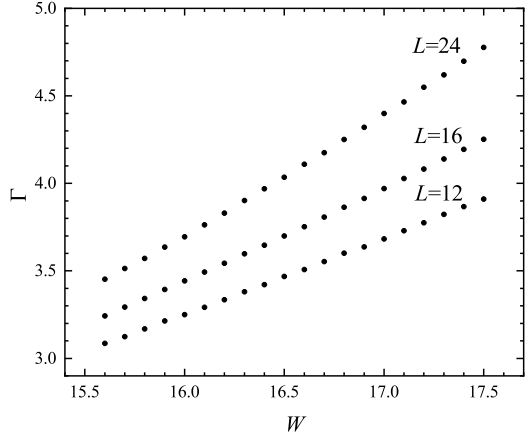}
\caption{Estimates of the quantity $ \Gamma =\left< \tilde{\gamma}_N \right> L$ for ensembles of $L \times L \times L$ cubes with
$L=12, 16$ and $24$ for disorder $W$ in the range $[15.5,17.5]$. The standard error of the data is much smaller than the symbol size. Here, the fixed initial matrix $Q_0$
given in Eq. (\ref{eq:Q0}) has been used.}
\label{f1}
\end{figure}

The alternative that we consider here is to simulate an ensemble of much shorter samples and consider an ensemble average.
For simplicity we consider cubes with $L_x=L$. The scaling hypothesis Eq. (\ref{eq:ScalingGeneral}) then becomes
\begin{equation}\label{eq:CubicScaling}
  \Gamma = \left\langle {\tilde \gamma_N } \right\rangle L  = f\left( {\frac{L}{\xi },\frac{L}{\xi },\frac{L}{\xi }} \right) = {f_{\mathrm{3D}}}\left( {\frac{L}{\xi }} \right) \;.
\end{equation}
It seems that we can now take full advantage of modern computers by simulating an ensemble of samples in parallel.
However, when attempting to analyse the numerical data we run into a difficulty.
In writing Eq. (\ref{eq:CubicScaling}) we have neglected the dependence on the initial matrix $Q_0$.
The importance of this matrix is immediately made clear by reference to Fig. \ref{f1} where
we show data for the sample mean of $\tilde{\gamma}_N$ obtained with
\begin{equation}\label{eq:Q0}
  Q_0 = \left( \begin{array}{c}1_N \\0_N\end{array}\right)\;.
\end{equation}
The data do not exhibit a common crossing point in the vicinity of the critical disorder $W_c \approx 16.5$.
While at first glance it might appear that the inclusion of an irrelevant correction might suffice
to restore a common crossing point, this is not the case; after further contemplation of Fig. \ref{f1} we see that the required correction would have to be relevant not irrelevant.

Fortunately, this problem is easily solved. Instead of a fixed $Q_0$, we make $Q_0$ a random matrix that is sampled from a probability distribution
that is invariant under convolution with the transfer matrix distribution, i.e. by generating $2N\times N$ random matrices with orthogonal columns
with a distribution that is invariant or stationary under the operation\cite{slevin04}
\begin{equation}
  Q'R = {M_x}Q \;.
\end{equation}
For such a distribution, we see from Eq. (33) of Ref. \citen{slevin14} that $\tilde{\gamma}_N$ becomes a sum of i.i.d. random variables, from which
it follows that the dependence of $\left<\tilde{\gamma}_N\right>$ on the length $L_x$ vanishes. In this case
\begin{equation}\label{eq:Equivalence}
  \left\langle {\tilde \gamma_N } \right\rangle  = \gamma_N \;.
\end{equation}
It immediately follows that
\begin{equation}
  {f_{\mathrm{Q1D}}} = {f_{\mathrm{3D}}}\;.
\end{equation}

The next question is how to generate such matrices.  We have found that the following procedure works well.
We start with $Q_0$ given by Eq. (\ref{eq:Q0}) and calculate
\begin{equation}
  Q'R = {M_q} \cdots {M_1}Q_0 \,.
\end{equation}
The matrix $R$ is then discarded and we set $Q_0=Q'$. This calculation is then repeated a total of, say, $r$ times.
Note that, for a given sequence of transfer matrices, the result of this calculation depends only on the total number of transfer matrix multiplications, i.e. on the product of $q$ and $r$, not on $q$ and $r$ separately.
For a given $L$, we have found that, when a sufficient number of randomizing multiplications are
performed, the distribution of $\tilde{\gamma}_N$ becomes independent of the number of such
multiplications.
We judge this by applying the Kolmogorov-Smirnov test to the resulting data for $\tilde{\gamma}_N$ with different numbers of randomizing multiplications.
For sufficiently large number of randomizing multiplications we find that the Kolmogorov-Smirnov test is unable to distinguish the distribution of $\tilde{\gamma}_N$ obtained.
When performing the Kolmogorov-Smirnov test we think it is important that the ensemble size used for the test match
that of the ensemble size used to accumulate data for finite size scaling
since deviations which are not apparent for small ensemble sizes may well be revealed for larger ensemble sizes.

\begin{figure}
\includegraphics[width=0.5\textwidth]{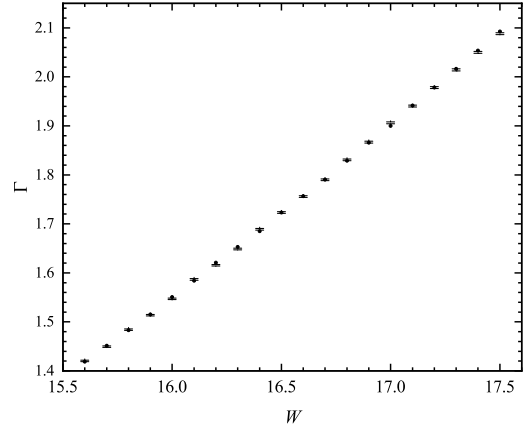}
\caption{A verification of Eq. (\ref{eq:Equivalence}) for $L$=12. Estimates of $\Gamma = \left< \tilde{\gamma}_N\right> L$ obtained using the parallel method (points without error bars)
are compared with estimates of $\Gamma = \gamma_N L$ obtained using the serial method (error bars without points).
For both sets of data the precision is approximately $0.1\%$.}
\label{f2}
\end{figure}

\begin{figure}
\includegraphics[width=0.5\textwidth]{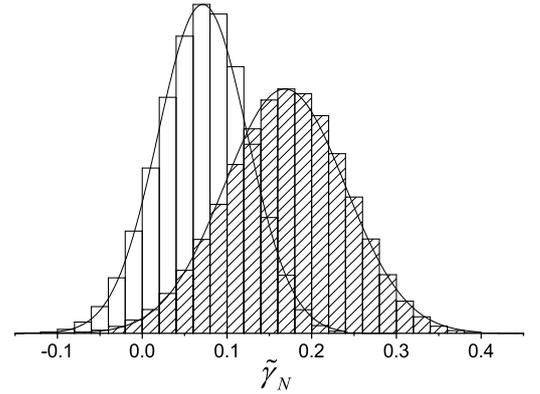}
\caption{Histograms for $L=24$ and $W=16.5$ of the distributions of $\tilde{\gamma}_N$ obtained
with random $Q_0$ (no shading) generated using 64 transfer matrix multiplications
and with fixed $Q_0$ (shaded) given by Eq. (\ref{eq:Q0}).
The ensemble size is 589824.}
\label{f3}
\end{figure}

We have verified  Eq. (\ref{eq:Equivalence}) for $L=12$ by comparing data obtained using the parallel method with data obtained using the serial method (see Fig. \ref{f2}).

To demonstrate further the importance of randomizing $Q_0$, we compare in Fig. \ref{f3}
the distribution of $\tilde{\gamma}_N$ obtained with fixed $Q_0$ given by Eq. (\ref{eq:Q0}) with the distribution
obtained with random $Q_0$ generated using 64 transfer matrix multiplications.
Note that, unlike eigenvalues whose order is arbitrary, the $\tilde{\gamma}_i$ are in
the order they are obtained in the QR factorization.
This is not in general in decreasing order. Nor is $\tilde{\gamma}_N$ always positive,
as is seen in Fig. \ref{f3} where the distributions extend to negative values.\cite{slevin04}
Another important point to grasp from Fig. \ref{f3} is that if too large an ensemble
with insufficient randomization of $Q_0$ is generated, the errors in the estimation
of  $\left< \tilde{\gamma}_N\right>$ will be systematic not random. This would make
reliable finite size scaling impossible.

Though not necessary for the estimation of the critical exponent, we can calculate all
$\tilde{\gamma}_1,\cdots \tilde{\gamma}_{2N}$  by making $Q_0$ a $2N \times 2N$ orthogonal matrix.
In this case, we have found that the $\tilde{\gamma}_i$ do not usually occur in pairs of opposite sign.
The only condition they satisfy in general is that their sum is zero.
Nevertheless, we have noticed that, if a sufficiently large number of transfer matrix multiplications is used to generate
random $Q_0$, the $\tilde{\gamma}_i$ again occur in pairs of opposite sign.\cite{slevin04}
This is similar to Eq. (\ref{eq:order}) but without the decreasing ordering.
However, this seems to require a much larger number of transfer matrix multiplications to generate random $Q_0$ than is needed when focussing as above only on
the distribution of $\tilde{\gamma}_N$.

\section{Numerical Simulation of Anderson's Model of Localisation}

\subsection{Details of the Simulations}

We have simulated ensembles of cubes with dimensions $L \times L \times L$ with
$L=12, 16, 24, 32, 48$ and $64$ and disorder $W$ in the range $[15.6,17.5]$.
For the largest system $L=64$ the range of disorder was restricted to $[15.7,17.3]$.

For each pair of $L$ and $W$ an ensemble of $589824$ samples was simulated and $\left< \tilde{\gamma}_N \right>$ estimated
using the sample mean with a precision given by the standard error in the mean calculated using the standard formulae.
In percentage terms, the precisions of the ensemble averages obtained varied between $0.07\%$
and $0.13\%$ depending on the pair of $W$ and $L$ considered.

To avoid round-off error, QR factorizations were performed after every 4 or 8 transfer matrix multiplications.

To obtain a stationary distribution of initial matrices $Q_0$, 64 transfer matrix multiplications
were used.
To check that this was sufficient we compared with data obtained with 32 and 96 multiplications. The
results of the Kolmogorov-Smirnov test for $L=48$ are shown in Table \ref{t1}.
It can be seen that, while 32 transfer matrix multiplications are not sufficient, the distributions
of $\tilde{\gamma}_N$ obtained with 64 and 96 multiplications cannot be distinguished with this number
of samples.
For $L=64$, a comparison of data obtained with 64 and 96 multiplications returned a p-value of 0.35 with the Kolmogorov-Smirnov test.

The computations were performed on the Supercomputer System B of the ISSP at the University of Tokyo.
Each calculation involved 288 MPI processes, with each process using 12 cores (with OpenMP).
The required parallel random number streams were generated using the MT2203 of Intel Math Kernel Library.
Since the calculation time scales as $L^7$, in practice virtually all the computer time is spent on the largest system size.

\begin{table*}[ht]
\caption{Example of the Kolmogorov-Smirnov test for cross section $48 \times 48$. The ensemble size is 589824.
The table shows the p-value returned by the Kolmogorov-Smirnov test. The rows and columns are labeled by the number of randomizing multiplications.}
\label{t1}
\begin{center}
\begin{tabular}{ccccc}
\hline
 \#multiplications & 32 & 64 & 96 \\
\hline
 32 & -& 0.004 & 0.023 \\
 64 & 0.004&- & 0.688 \\
 96 & 0.023& 0.688& -\\
\hline
\end{tabular}
\end{center}
\end{table*}

\section{Finite Size Scaling Analysis}

The critical disorder, critical exponent, and other critical quantities are estimated by fitting
the size and disorder dependence of the dimensionless quantity $\Gamma$ to a one parameter scaling model
\begin{equation}\label{eq:ScalingFunction}
  \Gamma  = \left< \tilde{\gamma}_N \right> L =  F\left( \phi  \right) \,.
\end{equation}
Here, $F$ is a scaling function and $\phi$ is a relevant scaling variable.
The scaling function is approximated by a Taylor series truncated at order $n$
\begin{equation}
  {F\left(\phi\right) = \sum\limits_{j = 0}^n {{F_j}{\phi ^j}} } \,.
\end{equation}
The scaling variable has the form
\begin{equation}\label{eq:phi}
  {\phi  = u\left( W - {W_c} \right){L^\alpha }} \,,
\end{equation}
where $\alpha$ is the inverse of the critical exponent
\begin{equation}
  {\nu  = \frac{1}{\alpha } > 0}\,,
\end{equation}
and $W_c$ is the critical disorder.
The function $u$ is approximated by a Taylor series truncated at order $m$,
\begin{equation}
  {u\left(  W - {W_c} \right) = \sum\limits_{j = 1}^m {{u_j}{ \left(W - {W_c}\right)^j}} }\,.
\end{equation}
To avoid any ambiguity in the model we impose the condition
\begin{equation}
  F_1=1 \,.
\end{equation}
The critical exponent $\nu$ is expected to be universal, i.e. it should be the same for all
Anderson transitions in three-dimensional systems in the orthogonal symmetry class.
The scaling function, and in particular the quantity
\begin{equation}
  \Gamma_c= F_0 = F\left(0\right)\,,
\end{equation}
are expected to be somewhat less universal, i.e. they should be the same for all
Anderson transitions in three-dimensional systems in the orthogonal symmetry class but depend
on the  boundary conditions imposed in the transverse directions\cite{Slevin00}.

To determine the best fit we perform a non-linear least squares fit, i.e. we minimize the $\chi^2$ statistic.
The quality of the fit is assessed using the goodness of fit probability $p$. Both the goodness of fit probability and the precision of the fitted parameters are determined by generating and fitting
an ensemble of 500 pseudo-data sets. The details of this procedure have already been described in
Ref. \citen{slevin14}.

\begin{figure}
\includegraphics[width=0.5\textwidth]{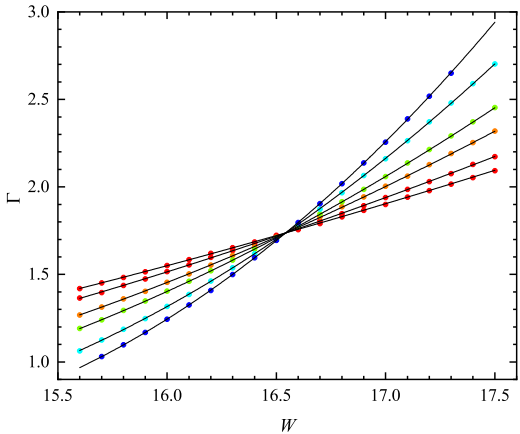}
\caption{(Color online)  The estimates (symbols) of $\Gamma = \left<\tilde{\gamma}_N\right>L$ for $L=$12, 16, 24, 32, 48 and $64$ for various disorders $W$ together with the finite size scaling fit (solid lines) described in the text.
The error bars of the numerical data are smaller than the symbol size.}
\label{f4}
\end{figure}

\begin{figure}
\includegraphics[width=0.5\textwidth]{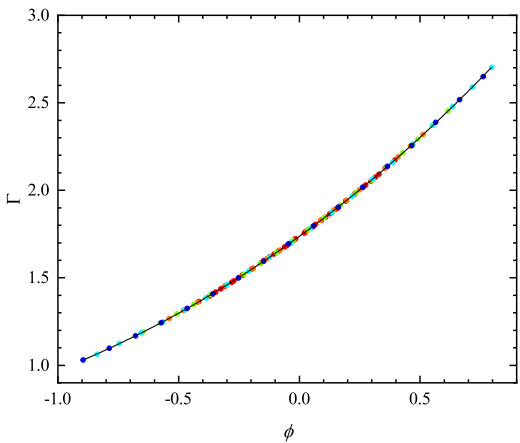}
\caption{ (Color online) The same data as in Fig. \ref{f4} but plotted versus the variable $\phi$ of Eq. (\ref{eq:phi}) to demonstrate the collapse of the data onto a common curve.
The line is the scaling function Eq. (\ref{eq:ScalingFunction}).}
\label{f5}
\end{figure}

In Fig. \ref{f4} we show a fit to 117 data points with $m=2$ and $n=3$. The orders of
the truncations are determined by requiring that the goodness of fit is greater than $0.1$ and that
the fit is reasonably stable against increases in both $m$ and $n$.
The details of the fit and values of the fitted parameters are shown in Table \ref{t2}.
To demonstrate one parameter scaling the collapse of all data points onto a single curve
when the data are re-plotted versus $\phi$ is shown in Fig. \ref{f5}.

\begin{table*}[ht]
\begin{center}
\begin{tabular}{l|llllll}
\hline
  & $\nu$ & $\Gamma_c$ & $W_c$ &  $N_D$  & $N_P$ & $p$ \\
\hline
 all data       & 1.572[1.566,1.577] & 1.7372[1.7359,1.7384]& 16.543[16.541,16.545] & 117 & 7 & 0.5\\
 restricted $W$ & 1.565[1.544,1.586] & 1.737[1.736,1.739]   & 16.542[16.540,16.545] & 48 & 6 & 0.6 \\
 restricted $L$ & 1.575[1.567,1.583] & 1.740[1.738,1.742]   & 16.546[16.543,16.549] & 77 & 7 & 0.8 \\
\hline
\end{tabular}
\caption{The details of the finite size scaling fits to all the data, to data with the range of
disorder $W$ restricted to [16.2,16.9], and to data with larger system sizes $L=24,32,48,64$ only.
The precisions are expressed as 95\% confidence intervals.
The values of the $\chi^2$ statistic for the best fits are 112.1, 38.5, and 59.8 respectively.}
\label{t2}
\end{center}
\end{table*}

\section{Discussion}

We have described an adaptation of the transfer matrix method often employed in the field of
Anderson localisation for massively parallel supercomputers.
We have illustrated the use of this method by applying it to Anderson's model of localisation
in three dimensions and estimated the critical exponent
\begin{equation}\label{eq:Result}
  \nu = 1.572 \pm 0.003 \,.
\end{equation}
(Note that the error here is a standard error not a 95\% confidence interval.)
The largest systems size considered here was $L=64$ and the precision of the critical exponent obtained  is approximately 0.2\%.
This is compared to the largest system size of $L=24$ in Ref. \citen{slevin14} where a precision of  0.35\% was obtained.
An important difference with Ref. \citen{slevin14} is that we did not need to consider corrections to scaling due to an irrelevant scaling variable.
The smallest transverse size used here is $L=12$.
As can be seen from Fig. 5 of Ref. \citen{slevin14}, irrelevant corrections are already less
than the precision of our data for $L\ge 12$.
Our estimate Eq. (\ref{eq:Result}) is consistent with that obtained by multi-fractal analysis of
eigenstates\cite{alberto10,alberto11,varga15} and with both numerical and experimental work
on the quantum kicked rotor.\cite{lemarie09,lemarie09b,lopez12}

In this work we simulated an ensemble of cubes, i.e. an ensemble of systems with aspect ratio fixed to unity.
However, this choice is not optimal.
In our calculation, about half the time was spent randomising the initial matrix $Q_0$ and about half the time estimating the ensemble average of $\tilde{\gamma}_N$.
A more economical approach would be to simulate a smaller ensemble of a longer systems.
More of the computer time would then be devoted to estimating the ensemble average of $\tilde{\gamma}_N$ rather than being ``wasted'' on randomising $Q_0$.
Indeed, provided the matrices $Q_0$ are randomised properly, we
see from equation Eq. (\ref{eq:Equivalence}) that there is no need to keep the aspect ratio fixed
and any convenient length can be simulated.
The appropriate choice will depend on the time limits set in the queuing system of the supercomputer being used.

In the serial method the length of the system is increased until the desired precision of the Lyapunov exponent is obtained.
Thus, the precise number of transfer matrix multiplications  is usually not known in advance.
This is inconvenient when we wish to simulate systems with correlated random potentials such as quantum Hall systems \cite{Ando84,Ando85,chalker88,kramer05} or cold atom systems\cite{delande14}.
The method we describe here may be better suited to such problems.

By allowing full exploitation of current supercomputers, the method described here may also be useful when studying higher dimensional systems\cite{garcia-garcia07,ueoka14,slevin16} where the time constraints of the transfer matrix method become more severe.

\begin{acknowledgment}

This work was supported by JSPS KAKENHI Grants No. 15H03700,  17K18763 and No. 26400393.
The authors thank the Supercomputer Center, the Institute for Solid State Physics, the University of Tokyo for the use of System B.
\end{acknowledgment}

\bibliography{Slevin2018JPSJ}

\begin{thebibliography}{10}

\bibitem{pichard81}
J.-L. Pichard and G.~Sarma: J. Phys. {\bfseries C14} (1981) L127.

\bibitem{mackinnon81}
A.~MacKinnon and B.~Kramer: Phys. Rev. Lett. {\bfseries 47} (1981) 1546.

\bibitem{MacKinnon83}
A.~MacKinnon and B.~Kramer: Z. Phys.B {\bfseries 53} (1983) 1.

\bibitem{abrahams79}
E.~Abrahams, P.~W. Anderson, D.~C. Licciardello, and T.~V. Ramakrishnan: Phys.
  Rev. Lett. {\bfseries 42} (1979) 673.

\bibitem{Kramer93}
B.~Kramer and A.~MacKinnon: Reports on Progress in Physics {\bfseries 56}
  (1993) 1469.

\bibitem{slevin97}
K.~Slevin and T.~Ohtsuki: Phys. Rev. Lett. {\bfseries 78} (1997) 4083.

\bibitem{slevin99}
K.~Slevin and T.~Ohtsuki: Phys. Rev. Lett. {\bfseries 82} (1999) 382.

\bibitem{slevin14}
K.~Slevin and T.~Ohtsuki: New Journal of Physics {\bfseries 16} (2014) 015012.

\bibitem{ueoka14}
Y.~Ueoka and K.~Slevin: Journal of the Physical Society of Japan {\bfseries 83}
  (2014) 084711.

\bibitem{slevin16}
K.~Slevin and T.~Ohtsuki: Journal of the Physical Society of Japan {\bfseries
  85} (2016) 104712.

\bibitem{asada02}
Y.~Asada, K.~Slevin, and T.~Ohtsuki: Phys. Rev. Lett. {\bfseries 89} (2002)
  256601.

\bibitem{Wigner51}
E.~P. Wigner: Mathematical Proceedings of the Cambridge Philosophical Society
  {\bfseries 47} (1951) 790.

\bibitem{Dyson61}
F.~J. Dyson: Journal of Mathematical Physics {\bfseries 3} (1962) 140.

\bibitem{Dyson62}
F.~J. Dyson: Journal of Mathematical Physics {\bfseries 3} (1962) 1199.

\bibitem{Huckestein90}
B.~Huckestein and B.~Kramer: Phys. Rev. Lett. {\bfseries 64} (1990) 1437.

\bibitem{slevin09}
K.~Slevin and T.~Ohtsuki: Phys. Rev. B {\bfseries 80} (2009) 041304.

\bibitem{amado11}
M.~Amado, A.~V. Malyshev, A.~Sedrakyan, and F.~Dom\'inguez-Adame: Phys. Rev.
  Lett. {\bfseries 107} (2011) 066402.

\bibitem{obuse12}
H.~Obuse, I.~A. Gruzberg, and F.~Evers: Phys. Rev. Lett. {\bfseries 109} (2012)
  206804.

\bibitem{anderson58}
P.~W. Anderson: Phys. Rev. {\bfseries 109} (1958) 1492.

\bibitem{Slevin00}
K.~Slevin, T.~Ohtsuki, and T.~Kawarabayashi: Phys. Rev. Lett. {\bfseries 84}
  (2000) 3915.

\bibitem{oseledec68}
V.~I. Oseledec: Trans. Moscow Math. Soc. {\bfseries {19}} (1968) 197.

\bibitem{slevin01a}
K.~Slevin and T.~Ohtsuki: Phys. Rev. B {\bfseries 63} (2001) 045108.

\bibitem{slevin04}
K.~Slevin, Y.~Asada, and L.~I. Deych: Phys. Rev. B {\bfseries 70} (2004)
  054201.

\bibitem{alberto10}
A.~Rodriguez, L.~J. Vasquez, K.~Slevin, and R.~A. R\"omer: Phys. Rev. Lett.
  {\bfseries 105} (2010) 046403.

\bibitem{alberto11}
A.~Rodriguez, L.~J. Vasquez, K.~Slevin, and R.~A. R\"omer: Phys. Rev. B
  {\bfseries 84} (2011) 134209.

\bibitem{varga15}
L.~Ujfalusi and I.~Varga: Physical Review B {\bfseries 91} (2015) 184206.

\bibitem{lemarie09}
G.~Lemari\'e, J.~Chab\'e, P.~Szriftgise, J.~C. Garreau, B.~Gr\'emaud, and
  D.~Delande: Phys. Rev. A {\bfseries 80} (2009) 043626.

\bibitem{lemarie09b}
G.~Lemari\'e, B.~Gr\'emaud, and D.~Delande: EPL (Europhysics Letters)
  {\bfseries 87} (2009) 37007.

\bibitem{lopez12}
M.~Lopez, J.-F. Cl\'ement, P.~Szriftgiser, J.~C. Garreau, and D.~Delande: Phys.
  Rev. Lett. {\bfseries 108} (2012) 095701.

\bibitem{Ando84}
T.~Ando: Journal of the Physical Society of Japan {\bfseries 53} (1984) 3101.

\bibitem{Ando85}
T.~Ando and H.~Aoki: Journal of the Physical Society of Japan {\bfseries 54}
  (1985) 2238.

\bibitem{chalker88}
J.~Chalker and P.~Coddington: J. Phys. {\bfseries C21} (1988) 2665.

\bibitem{kramer05}
B.~Kramer, T.~Ohtsuki, and S.~Kettemann: Phys. Rep. {\bfseries 417} (2005) 211.

\bibitem{delande14}
D.~Delande and G.~Orso: Phys. Rev. Lett. {\bfseries 113} (2014) 060601.

\bibitem{garcia-garcia07}
A.~M. García-García and E.~Cuevas: Physical Review B {\bfseries 75} (2007)
  174203.

\end{thebibliography}

\end{document}